\documentclass[preprint,fleqn,showpacs,showkeys]{revtex4}
\usepackage{graphicx}
\usepackage{amssymb}
\usepackage{amsmath}
\usepackage{bm}
\usepackage{feynmf}
\usepackage{tikz}
\usetikzlibrary{trees}
\usetikzlibrary{decorations.pathmorphing}
\usetikzlibrary{decorations.markings}
\usetikzlibrary{snakes}

\def\be{\begin{equation}}
\def\ee{\end{equation}}
\def\bea{\begin{eqnarray}}
\def\eea{\end{eqnarray}}

\begin{document}
\setcounter{page}{1}
\title{A novel way of calculating scattering integrals}
\author{Alfredo Takashi \surname{Suzuki}}
\affiliation{Department of Physics, La Sierra University\\
4500 Riverwalk Pkw., Riverside, CA 92505}
\author{Timothy \surname{Suzuki}}
\affiliation{Department of Physics and Astronomy, Michigan State University\\
567 Wilson Rd., East Lansing, MI 48824}
\date[]{}

\begin{abstract}
The technique coined as NDIM - Negative Dimensional Integration Method by their discoverers, relies on a three-pronged basis: Gaussian integration, series expansion and analytic continuation. The technique has been successfully applied to the calculation of covariant and non covariant Feynman integrals in a generic dimensional regularization space, i.e., $D$-dimensional space-time for $D$ including the negative domain values. Since the dimensionality is general, we can use specifically for one-dimensional integrals. In this work we show how this technique can be applied to tackle certain improper integrals and give an example of a particular improper integral that appears in quantum mechanical scattering process.  Traditionally, improper integrals are ascribed certain values through the limiting approach or as is known, by the Cauchy principal value via residues concept technique. Here we use the NDIM approach to do the calculations and show it works fine for the improper integrals. This novel approach we believe is more straightforward and does not require to handle poles, residues, or difficult closed contours as in the traditional approach. 
\end{abstract}



\maketitle

\section{Introduction}

\vspace{.5cm}

NDIM (Negative Dimensional Integration Method) \cite{Halliday-Ricotta} was born within the context of perturbative quantum field theories where Feynman diagrams that give a pictorial view of physical processes are calculated. When the perturbative series entails loop calculations, these often lead to diverging Feynman integrals, for which a regularization procedure is called for. Among several regularization procedures, dimensional regularization \cite{DR} is a popular choice, where a four-dimensional integration measure is analytically extended to a $D$-dimensional one. In essence, what NDIM does is to allow this analytic continuation to span even into the negative dimensionality region. With the help of a $D$-dimensional Gaussian integration and series expansion, an integrand with poles are transformed into polynomials that are integrated in negative dimensions. After the polynomial integration in negative dimensions is carried out, analytic continue the result back into positive dimensions. \cite{SuzukiQFT}.

This $D$-dimensional NDIM procedure can be recast into a one-dimensional integration measure without difficulty \cite{Suzuki, Suzuki2}. In our previous works, we considered integrands with pure imaginary poles, so the calculation entails no closed circuit with real poles standing on the integration path. In the present work we consider improper integrals with real poles, and show that even for this case NDIM can be applied successfully.

\section{Improper integrals}

Let us consider the following improper integral:
\bea
\label{Integral}
I  & = &   \int\limits_{-\infty}^{+\infty}  \frac{dx}{\left(x^2-a^2\right)}=\int\limits_{-\infty}^{+\infty}  \frac{dx}{\left(x-a\right)\left(x+a\right)}\,;\quad a\in {\mathbb R}.
\eea

Since the integrand is not well-defined (diverges) at $x=\pm a$, we may try to ascribe a value to this integral by the limiting procedure as follows:
\bea
\label{Integral2}
I  & = &   \frac{1}{2a}\left\{ \lim_{\stackrel{\epsilon\to 0^+;}{R\to \infty}}\left[ \int\limits_{-R}^{a-\epsilon}  \frac{dx}{\left(x-a\right)}+\int\limits_{a+\epsilon}^{+R} \frac{dx}{\left(x-a\right)}\right] \right. \nonumber\\
   &  & \hspace {0.5cm}  + \left. { \lim_{\stackrel{\epsilon^\prime\to 0^+;}{R\to \infty}}}\left[ \int\limits_{-R}^{-a-\epsilon^\prime}  \frac{dx}{\left(x+a\right)}+\int\limits_{-a+\epsilon^\prime}^{+R} \frac{dx}{\left(x+a\right)}\right] \right\}.
\eea

We can ascribe a principal value (PV) to it in the sense of Cauchy by using the same infinitesimal, that is, $\epsilon=\epsilon^\prime \to 0^+$, in the limiting analysis
\bea
\label{PV}
I_{\rm PV}  & = &    \frac{1}{2a}\left\{ \lim_{\stackrel{\epsilon\to 0^+;}{R\to \infty}}\left[ \int\limits_{-R}^{a-\epsilon}  \frac{dx}{\left(x-a\right)}+\int\limits_{a+\epsilon}^{+R} \frac{dx}{\left(x-a\right)}\right] \right. \nonumber\\
   &  & \hspace {0.5cm}  + \left. { \lim_{\stackrel{\epsilon\to 0^+;}{R\to \infty}}}\left[ \int\limits_{-R}^{-a-\epsilon}  \frac{dx}{\left(x+a\right)}+\int\limits_{-a+\epsilon}^{+R} \frac{dx}{\left(x+a\right)}\right] \right\} = 0.
\eea

We could also use the closed contour integration analytically continuing the integration variable into complex variables, $x \to z$, 
\bea
\label{ComplexIntegral}
I_{\rm PV}  & = &   \oint  \frac{dz}{\left(z^2-a^2\right)}=\oint  \frac{dz}{\left(z-a\right)\left(z+a\right)}\,;\quad z\in {\mathbb C}; \:\:a\in {\mathbb R}.
\eea

Here, because we have two poles on the real axis, $x=\pm a$, we need to distort the contour around them by a semi-circle of infinitesimal radius ($r \to 0^+$) in order to avoid the poles. We can do this in several manners, closing the contour either on the upper or lower hemispheres:

1. Excluding both poles;

2. Including both poles;

3. Excluding the pole $x=-a$ and including the pole $x=a$;

4. Including the pole $x=-a$ and excluding the pole $x=a$.

All theses choices are totally equivalent, so, if we choose, for instance, the following contour, excluding both poles from the closed upper hemisphere circuit:

\begin{tikzpicture}
\draw [->,densely dotted] (-3.7,0) -- (3.7,0);
\draw (4.2,0) node {${\mathfrak {Re}} z$}; 
\draw [->, densely dotted] (0,-0.5) -- (0,3.7);
\draw (0,4.0) node {${\mathfrak {Im}} z$};
\draw [blue]  [->] (-3.0,0) -- (-2.55,0);
\draw [blue] [-](-2.55,0) -- (-2.2,0);
\draw [blue][->] (-1.2,0) -- (0,0);
\draw [blue] [->] (-2.2,0) arc [start angle=180, end angle=90, radius=0.5];
\draw [blue] [-] (-1.7,0.5) arc [start angle=90, end angle=0, radius=0.5];
\draw  [blue] [-] (0,0) -- (1.2,0);
\draw [blue] [->] (1.2,0) arc [start angle=180, end angle=90, radius=0.5];
\draw [blue] [-] (1.7,0.5) arc [start angle=90, end angle=0, radius=0.5];
\draw[blue] [->](2.2,0) -- (2.65,0);
\draw[blue] (2.65,0) -- (3.0,0);
\draw [blue] [->] (3.0,0.0) arc [start angle=0, end angle=45, radius=3.0];
\draw [blue] [->] (2.12,2.12) arc [start angle=45, end angle=90, radius=3.0];
\draw [blue] [->] (0,3.0) arc [start angle=90, end angle=135, radius=3.0];
\draw [blue] (-2.12,2.12) arc [start angle=135, end angle=180, radius=3];
\draw (-3.3,0) node [below=0.1pt]  {$-R$};
\draw (3.2,0) node [below=0.1pt] {$+R$};
\draw (2.3,2) node [above=1pt] {$C_{\!R}$};
\draw (-2.3,2.1) node [above=1pt] {$C_{\!R}$};
\draw (-1.7,0.4) node [above=1pt] {$C_{\!r}^-$};
\draw (1.7,0.4) node [above=1pt] {$C_{\!r}^+$};
\filldraw[red] (-1.7,0.02) node {$\star$};
\draw[red] (-1.7,0.3) node {\small $-a$};
\filldraw[red] (1.7,0.02) node {$\star$};
\draw[red] (1.7,0.3) node {\small $a$};
\draw (0,0) node [below=15pt]  {Figure 1};
\end{tikzpicture}

Cauchy residue theorem yields this closed loop integration as, with $f(z) \equiv (z^2-a^2)^{-1}=[(z-a)(z+a)]^{-1}$
\bea
\label{Cauchy}
I_{\rm PV}  & = &   \oint  \frac{dz}{\left(z^2-a^2\right)}=2i\pi \sum {\rm Residues} f(z) = 0,
\eea
since the chosen closed contour does not contain any poles. 

The contour integration, on the other hand, can be broken down into
\bea
\label{Cauchypieces}
I_{\rm PV}  & = &   \oint  f(z)dz \nonumber\\
                  & = & \int_{C_R} f(z)dz +\int\limits_{-R}^{-a-r}f(z)dz+\int_{C_r^-} f(z)dz\nonumber\\
                  & + & \int\limits_{-a+r}^{a-r}f(z)dz+\int_{C_r^+} f(z)dz+\int\limits_{a+r}^{+R}f(z)dz     
\eea

Let us look at in detail these integrals. First, let us consider
\be
C_R \equiv \int_{C_R} f(z)dz = \int_{C_R} \frac{dz}{(z-a)(z+a)}.
\ee 

For this integral, along the upper hemisphere counterclockwise semi-circle, we have
\bea 
z & = & Re^{i\theta}\,,\quad  0\le \theta \le \pi; \nonumber \\
dz & = & iRe^{i\theta} d\theta.
\eea

Therefore,
 \be
C_R = \int_{0}^\pi \frac{iRe^{i\theta}d\theta}{(Re^{i\theta}-a)(Re^{i\theta}+a)}.
\ee 

For the relevant limit $R\to \infty$, the integrand's leading term is 
\bea
\lim_{R\to \infty} C_R & \approx &  \frac{i}{R}\int_{0}^\pi e^{-i\theta}d\theta \to 0.
\eea 

Next, let us examine
\be
C_r^- \equiv \int_{C_r^-} f(z)dz = \int_{C_r^-} \frac{dz}{(z-a)(z+a)}.
\ee 

For this integral, along the upper hemisphere clockwise semi-circle around $x=-a$, we have
\bea 
z & = & -a+re^{i\varphi}\,,\quad  \pi\le \varphi \le 0; \nonumber \\
dz & = & ire^{i\varphi} d\varphi.
\eea

Therefore,
\bea
C_r^- & = &  \int_{\pi}^0 \frac{ire^{i\varphi}d\varphi}{(-2a+re^{i\varphi})(re^{i\varphi})}\nonumber\\
          & = & i\int_\pi^0 \frac{d\varphi}{(-2a+re^{i\varphi})}.
\eea 

Here, for the relevant limit $r \to 0^+$, the integrand's leading term is
\bea
\lim_{r\to 0^+} C_r^- & \approx &  -\frac{i}{2a}\int_\pi^0 {d\varphi}\to \frac{i\pi}{2a}.
\eea 

Now, for the 
\be
C_r^+ \equiv \int_{C_r^+} f(z)dz = \int_{C_r^+} \frac{dz}{(z-a)(z+a)}.
\ee 

For this integral, along the upper hemisphere clockwise semi-circle around $x=a$, we have
\bea 
z & = & a+re^{i\varphi}\,,\quad  \pi\le \varphi \le 0; \nonumber \\
dz & = & ire^{i\varphi} d\varphi.
\eea

Therefore,
\bea
C_r^+ & = &  \int_{\pi}^0 \frac{ire^{i\varphi}d\varphi}{(re^{i\varphi})(2a+re^{i\varphi})}\nonumber\\
          & = & i\int_\pi^0 \frac{d\varphi}{(2a+re^{i\varphi})}.
\eea 

Here, for the relevant limit $r \to 0^+$, the integrand's leading term is 
\bea
\lim_{r\to 0^+} C_r^+ & \approx &  \frac{i}{2a}\int_\pi^0 {d\varphi}\to \frac{-i\pi}{2a}.
\eea 

Finally, plugging the results for $C_R$, $C_r^-$ and $C_r^+$ into equation (\ref{Cauchypieces}) with the relevant limits taken, $R\to \infty; r\to 0^+$, we have
\bea
\label{Cauchypv}
I_{\rm PV}  & = &   \oint  f(z)dz \nonumber\\
                  & = & 0 +\int\limits_{-\infty}^{-a}f(x)dx+\frac{i\pi}{2a}+\int\limits_{-a}^{a}f(x)dx-\frac{i\pi}{2a}+\int\limits_{a}^{+\infty}f(x)dx    \nonumber\\
                  & = &  \int\limits_{-\infty}^{+\infty}f(x)dx\equiv  \int\limits_{-\infty}^{+\infty}\frac{dx}{(x^2-a^2)}.
\eea

Comparing with equation (\ref{Cauchy}), we get, in accordance with equation (\ref{PV})
\bea
\int\limits_{-\infty}^{+\infty}\frac{dx}{(x^2-a^2)}=0.
\eea

Since the integral in equation (\ref{Integral}) is an improper one, we can in principle ascribe other values to it, other than the PV value just calculated. It is possible to consider the original integral, equation (\ref{Integral}), with $a \in {\mathbb C}$, by letting the real pole be dislocated by an infinitesimal imaginary part, i.e., either $a_+ = a+i\varepsilon$ or $a_- =a-i\varepsilon$.

\subsection{Improper integral for $a_+$ shift}
Let us consider now the following integral,
\bea
\label{+}
I_+=\lim_{\varepsilon \to 0^+}I_{+}(\varepsilon),
\eea
where
\bea
I_{+} (\varepsilon) & \equiv & \int\limits_{-\infty}^{+\infty}  \frac{dz}{\left(z^2-a_+^2\right)}\,;\quad a_+\in {\mathbb C}\nonumber\\
         & = & \int\limits_{-\infty}^{+\infty}  \frac{dz}{\left(z-a-i\varepsilon\right)\left(z+a+i\varepsilon\right)}\,, \quad a\in {\mathbb R}.
\eea

For this, we consider the following closed circuit,

\begin{tikzpicture}
\draw [->,densely dotted] (-3.7,0) -- (3.7,0);
\draw (4.2,0) node {${\mathfrak {Re}} z$}; 
\draw [->, densely dotted] (0,-1.0) -- (0,3.7);
\draw (0,4.0) node {${\mathfrak {Im}} z$};
\draw [blue]  [->] (-3.0,0) -- (-1.5,0);
\draw [blue] (-1.5,0) -- (0,0);
\draw  [blue] [->] (0,0) -- (1.5,0);
\draw[blue] (1.5,0) -- (3.0,0);
\draw [blue] [->] (3.0,0.0) arc [start angle=0, end angle=45, radius=3.0];
\draw [blue] [->] (2.12,2.12) arc [start angle=45, end angle=90, radius=3.0];
\draw [blue] [->] (0,3.0) arc [start angle=90, end angle=135, radius=3.0];
\draw [blue] (-2.12,2.12) arc [start angle=135, end angle=180, radius=3];
\draw (-3,0) node [below=0.1pt]  {$-R$};
\draw (3,0) node [below=0.1pt] {$+R$};
\draw (2.3,2) node [above=1pt] {$C_{\!R}$};
\draw (-2.3,2.1) node [above=1pt] {$C_{\!R}$};
\filldraw[red] (-1.7,-0.30) node {$\star$};
\draw[red] (-2.45,-0.50) rectangle (-0.95,-1.0);
\draw[red] (-1.7,-0.75) node {\small $-a-\!i\varepsilon$};
\filldraw[red] (1.7,+0.30) node {$\star$};
\draw[red] (1.05,0.50) rectangle (2.3,1.0);
\draw[red] (1.7,0.75) node {\small $a+\!i\varepsilon$};
\draw (0,0) node [below=30pt]  {Figure 2};
\end{tikzpicture}

Then, we have
\bea
I_{+}(\varepsilon) & \equiv & \oint  \frac{dz}{\left(z^2-a_+^2\right)} = \oint  \frac{dz}{\left(z-a-i\varepsilon\right)\left(z+a+i\varepsilon\right)} \nonumber\\
         & = & 2i\pi \sum {\rm Residues} f_+(z)\,, \quad f_+(z)\equiv \frac{1}{\left(z-a-i\varepsilon\right)\left(z+a+i\varepsilon\right)}\nonumber \\
         & = & 2i\pi \left\{ \lim_{z \to a+i\varepsilon }(z-a-i\varepsilon)f_+(z) \right\}\nonumber\\
         & = & \frac{i\pi}{a+i\varepsilon}.
\eea

So, remembering that the large counterclockwise contour does not contribute, i.e.,  $C_R \to 0$ for $R\to \infty$, we have 
\be
I_+ = \lim_{\varepsilon \to 0^+}\frac{i\pi}{a+i\varepsilon} = \frac{i\pi}{a} 
\ee

\subsection{Improper integral for $a_-$ shift}
For this integral, we have
\bea
\label{-}
I_-=\lim_{\varepsilon \to 0^+}I_{-}(\varepsilon),
\eea
where
\bea
I_{-}(\varepsilon) & \equiv & \int\limits_{-\infty}^{+\infty}  \frac{dz}{\left(z^2-a_-^2\right)}\,;\quad a_-\in {\mathbb C}\nonumber\\
         & = & \int\limits_{-\infty}^{+\infty}  \frac{dz}{\left(z-a+i\varepsilon\right)\left(z+a-i\varepsilon\right)}\,, \quad a\in {\mathbb R}.
\eea

The same previous closed circuit now encloses the pole in the second quadrant,

\begin{tikzpicture}
\draw [->,densely dotted] (-3.7,0) -- (3.7,0);
\draw (4.2,0) node {${\mathfrak {Re}} z$}; 
\draw [->, densely dotted] (0,-1.0) -- (0,3.7);
\draw (0,4.0) node {${\mathfrak {Im}} z$};
\draw [blue]  [->] (-3.0,0) -- (-1.5,0);
\draw [blue] (-1.5,0) -- (0,0);
\draw  [blue] [->] (0,0) -- (1.5,0);
\draw[blue] (1.5,0) -- (3.0,0);
\draw [blue] [->] (3.0,0.0) arc [start angle=0, end angle=45, radius=3.0];
\draw [blue] [->] (2.12,2.12) arc [start angle=45, end angle=90, radius=3.0];
\draw [blue] [->] (0,3.0) arc [start angle=90, end angle=135, radius=3.0];
\draw [blue] (-2.12,2.12) arc [start angle=135, end angle=180, radius=3];
\draw (-3,0) node [below=0.1pt]  {$-R$};
\draw (3,0) node [below=0.1pt] {$+R$};
\draw (2.3,2) node [above=1pt] {$C_{\!R}$};
\draw (-2.3,2.1) node [above=1pt] {$C_{\!R}$};
\filldraw[red] (-1.7,0.30) node {$\star$};
\draw[red] (-2.45,0.50) rectangle (-0.95,1.0);
\draw[red] (-1.7,0.75) node {\small $-a+\!i\varepsilon$};
\filldraw[red] (1.7,-0.30) node {$\star$};
\draw[red] (1.05,-0.50) rectangle (2.3,-1.0);
\draw[red] (1.7,-0.75) node {\small $a-\!i\varepsilon$};
\draw (0,0) node [below=30pt]  {Figure 3};
\end{tikzpicture}

Then, we have
\bea
I_{-} (\varepsilon) & \equiv & \oint  \frac{dz}{\left(z^2-a_-^2\right)} = \oint  \frac{dz}{\left(z-a+i\varepsilon\right)\left(z+a-i\varepsilon\right)} \nonumber\\
         & = & 2i\pi \sum {\rm Residues} f_-(z)\,, \quad f_-(z)\equiv \frac{1}{\left(z-a+i\varepsilon\right)\left(z+a-i\varepsilon\right)}\nonumber \\
         & = & 2i\pi \left\{ \lim_{z \to -a+i\varepsilon }(z+a-i\varepsilon)f_-(z) \right\}\nonumber\\
         & = & \frac{i\pi}{-a+i\varepsilon}.
\eea

So, finally,
\be
I_- = \lim_{\varepsilon \to 0^+}\frac{i\pi}{-a+i\varepsilon} = -\frac{i\pi}{a} .
\ee

We note that, as the PV is also defined by 
\bea
I_{\rm PV} = \frac{1}{2}\lim_{\varepsilon \to 0^+}\left\{ I_+(\varepsilon)+I_-(\varepsilon)\right\} =0.
\eea

\section{NDIM and the improper integrals}

NDIM has already been applied for some definite integrals \cite{Suzuki, Suzuki2}. Here we are going to employ the NDIM to evaluate equation (\ref{Integral}).
To implement the NDIM technique for the sought integral, let us introduce the generating functional Gaussian integral that is pertinent to our calculation:
\bea
\label{Gaussian_generating}
G_a & = & \int\limits_{-\infty}^{+\infty}  dx \;e^{-\lambda(x^2-a^2)}. 
\eea

The Gaussian integration can be performed without difficulty, yielding
\bea
G_a & = &  e^{\lambda a^2}\!\!\!\int\limits_{-\infty}^{+\infty} e^{-\lambda x^2} \nonumber\\
                    & = & e^{\lambda a^2} \sqrt{\frac{\pi}{\lambda}} .
                    \eea

Next, expanding the exponential function in the result above in power series we get
\bea
\label{Gaussian_result}
G_a & = & \pi^{1/2}\sum_{n=0}^{\infty}  \frac{a^{2n}}{n!}\lambda^{n-1/2}.
\eea

On the other hand, expanding in power series the original Gaussian integral (\ref{Gaussian_generating}), we have
\bea
\label{Gaussian_NDIM}
G_a & = &  \int\limits_{-\infty}^{+\infty} dx \sum_{k=0}^{\infty}(-1)^k\frac{\lambda^k}{k!}  (x^2-a^2)^k.
\eea

As the above integral has a polynomial integrand equivalent to the desired integral with poles, we have the negative dimensional integration characterized by this procedure, i.e,
introducing the notation 
\bea
\label{I_NDIM}
I_{\rm NDIM}(k) & = & \int \limits_{-\infty}^{+\infty} \hat{d}x\left(x^2-a^2\right)^k,
\eea
with $\hat{d}x$ to remind that we are in the negative dimension measure continuation. 
Then, the later (\ref{Gaussian_NDIM}) can be written as 
\bea
\label{Def}
G_a & = &  \sum_{k=0}^{\infty}(-1)^k\frac{\lambda^k}{k!} I_{\rm NDIM}(k).
\eea

Comparing the two series expansion for $G_a$, term by term, we have that $n=k+1/2$ and $I_{\rm NDIM}(k)$ can be obtained as
\bea
\label{NDIM result}
I_{\rm NDIM}(k) & = & (-1)^{-k} k! \pi^{1/2} \frac{a^{2k+1}}{(k+1/2)!} \nonumber\\
                          & = & (-1)^{-k} \pi^{1/2}a^{2k+1}   \frac{\Gamma(1+k)}{\Gamma(1+k+1/2)} \nonumber\\
                          & = & (-1)^{-k} \pi^{1/2}a^{2k+1}   \frac{1}{(1+k)_{1/2}},
\eea
where in the last line above we have introduced the Pochhammer's symbol 
\be
(\alpha)_\beta \equiv \frac{\Gamma(\alpha+\beta)}{\Gamma(\alpha)}.
\ee

Since we want (\ref{I_NDIM}) analytically continued (AC) to allow for negative values of $k$ (in positive dimensional measure) we proceed by using in (\ref{NDIM result}) the following Pochhammer's identity
\be
(1-\alpha)_{\beta}=(-1)^{\beta}\frac{1}{(\alpha)_{-\beta}}.
\ee

Then,
\bea
\label{AC}
I_{\rm NDIM}^{\rm AC}(k) & = & (-1)^{-k-1/2} \pi^{1/2}a^{2k+1}  (-k)_{-1/2},
\eea

In (\ref{I_NDIM}), we want $k=-1$ to reproduce the original integral in positive dimensions, so we are interested in
\bea
\label{AC}
I_{\rm NDIM}^{\rm AC}(k=-1) & = & (-1)^{1/2} \pi^{1/2}a^{-1}  (1)_{-1/2}\nonumber\\
                                               & = & \frac{i\pi}{a} = I_+.
\eea

Since the original integral is, of course invariant under the symmetry $a \to -a$, it follows that we also have
\bea
\label{AC}
I_{\rm NDIM}^{\rm AC}(k=-1) & = & (-1)^{1/2} \pi^{1/2}(-a)^{-1}  (1)_{-1/2}\nonumber\\
                                               & = & -\frac{i\pi}{a} = I_-.
\eea

And once again, we can calculate the PV value,
\bea
I_{\rm PV} = \frac{1}{2}\lim_{\varepsilon \to 0^+}\left\{ I_+(\varepsilon)+I_-(\varepsilon)\right\} =0.
\eea

\section{NDIM in quantum mechanics scattering calculation}

In quantum mechanics scattering problems, there is an improper integral of the following form to be calculated:
\bea
\label{S}
S(\sigma) & =  & \int\limits_{-\infty}^{+\infty} dx \frac{ x \sin x}{(x^2-\sigma^2) }, \quad \sigma \in {\mathbb R}
\eea

In order to perform this integration using NDIM, we first need to express it in terms of integrands that would be fitting for applying the technique. So the `road preparation' for it is done by considering the integral
\bea
\label{Exponential}
E(a, \sigma) & = & \int\limits_{-\infty}^{+\infty} dx \frac{e^{iax}}{(x^2-\sigma^2)}\,,\quad a\in{\mathbb R}.\\
                    & = & \sum_{m=0}^{\infty}\frac{(ia)^m}{m!}I^m(\sigma), \nonumber
\eea
where we have defined
\be
\label{Im}
I^m(\sigma) = \int\limits_{-\infty}^{+\infty}dx \frac{x^m}{(x^2-\sigma^2)}.
\ee

We introduce then the relevant generating functional Gaussian integration as follows
\bea
G(\alpha,\beta) & = & \int\limits_{-\infty}^{+\infty}dx e^{\alpha x-\beta(x^2-\sigma^2)} \label{G}\\
                         & = & e^{\frac{\alpha^2}{4\beta}+\beta \sigma^2}\sqrt{\frac{\pi}{\beta}}\, \nonumber\\
                         & = & \sqrt{\frac{\pi}{\beta}}\,\sum_{k,l=0}^{\infty} \frac{\left(\beta\sigma^2\right)^k}{k!}\frac{\alpha^{2l}}{4^l \beta^l l!}.\label{G1}
\eea

On the other hand, series expansion of integrand in (\ref{G}) gives
\bea
G(\alpha,\beta) & = & \sum_{r,s=0}^{\infty} (-1)^s \frac{\alpha^r \beta^s}{r!s!}\int\limits_{-\infty}^{+\infty}dx\,x^r \left(x^2-\sigma^2\right)^s \label{G2}.
\eea

Comparing (\ref{G1}) and (\ref{G2}) we have that
\bea
 \sum_{r,s=0}^{\infty} (-1)^s \frac{\alpha^r \beta^s}{r!s!}\int\limits_{-\infty}^{+\infty}dx\,x^r \left(x^2-\sigma^2\right)^s=\pi^{1/2}\,\sum_{k,l=0}^{\infty} \frac{\sigma^{2k}}{4^l}\frac{\beta^{k-l-1/2}}{k!}\frac{\alpha^{2l}}{l!}.
\eea

The term by term equality entails 
\bea
r & = & 2l \nonumber\\
s & = & k-l-1/2.
\eea

Then, we have that $k=s+r/2+1/2$ and $l=r/2$, leading to
\bea 
\int\limits_{-\infty}^{+\infty}dx\,x^r \left(x^2-\sigma^2\right)^s & = & (-1)^{-s} r!s!\pi^{1/2}\frac{\beta^{2s+r+1}}{4^{r/2}}\frac{1}{(s+r/2+1/2)!}\frac{1}{(r/2)!}\nonumber\\
                                                                                              & = & (-1)^{-s} \pi^{1/2} \frac{\sigma^{r+2s+1}}{4^{r/2}} \frac{(1+r/2)_{r/2}}{(1+s)_{r/2+1/2}} \label{NDIM}.
\eea

Note that the integral
\be
I_{\rm NDIM}(r,s) \equiv \int\limits_{-\infty}^{+\infty}dx\,x^r \left(x^2-\sigma^2\right)^s,  \label{NDIMr}
\ee
will be equivalent to (\ref{Im}) when $r = m$ and $s = -1$. Note also that henceforth we are going to be omitting the reminder sign over $\hat dx$ for the NDIM integration measure. Since we need $s$ to assume negative values in $I_{\rm NDIM}(r,s) $, we analytic continue the Pochhammer's symbol on the righthandside of (\ref{NDIM}) that contains $s$, i.e.,
\be
\frac{1}{(1+s)_{r/2+1/2}} \stackrel{\rm AC}{\longrightarrow} \frac{(-s)_{-r/2-1/2}}{(-1)^{r/2+1/2}}.
\ee 

Then,
\be
I_{\rm NDIM}^{\rm AC}(r,s) =  (-1)^{-s-r/2-1/2} \pi^{1/2} \frac{\sigma^{r+2s+1}}{4^{r/2}} (1+r/2)_{r/2}(-s)_{-r/2-1/2}.
\ee

Comparing (\ref{Im}) with (\ref{NDIMr}) we see that we need to have $r=m$ and $s=-1$ in (\ref{NDIMr}) to get the relevant result for (\ref{Im}). So
\bea
I_{\rm NDIM}^{\rm AC}(m,-1) & =  & (-1)^{1/2-m/2} \pi^{1/2} \frac{\sigma^{m-1}}{4^{m/2}} (1+m/2)_{m/2}(1)_{-m/2-1/2}\nonumber\\
                                               & = & i\pi^{1/2}\frac{\sigma^m}{\sigma}\frac{1}{(-4)^{m/2}}\frac{\Gamma(1+m)\Gamma(1/2-m/2)}{\Gamma(1+m/2)}.
\eea

Now, using the gamma function relation
\be
\Gamma(1/2-m/2) = \pi^{1/2}(-4)^{m/2}\frac{\Gamma(1+m/2)}{\Gamma(1+m)}.
\ee

Then
\be
I_{\rm NDIM}^{\rm AC}(m,-1) \equiv  I^m(\sigma) = \frac{i\pi \sigma^m}{\sigma}.
\ee

Substituting this last result into (\ref{Exponential}) we get
\bea
\label{Er}
E(a, \sigma) & = & \sum_{m=0}^{\infty}\frac{(ia)^m}{m!}\frac{i\pi \sigma^m}{\sigma}\nonumber\\
                     & = & \frac{i\pi}{\sigma}\sum_{m=0}^{\infty}\frac{(ia\sigma)^m}{m!} .
\eea

Thus, finally:
\bea
\label{Eresult}
\int\limits_{-\infty}^{+\infty} dx \frac{e^{iax}}{(x^2-\sigma^2)} & = &  \frac{i\pi }{\sigma}\,e^{ia\sigma}.
\eea

Next, we need to evaluate
\be
\label{x}
E_x(a,\sigma) = \int\limits_{-\infty}^{+\infty} dx \frac{x\,e^{iax}}{(x^2-\sigma^2)}.
\ee

We do not need to recalculate from scratch; we use the identity
\be
x\,e^{iax} = -i \frac{d}{da}e^{iax},
\ee
in the result (\ref{Eresult}). Doing this, we get
\be
\label{xr}
E_x(a,\sigma) \equiv \int\limits_{-\infty}^{+\infty} dx \frac{x\,e^{iax}}{(x^2-\sigma^2)} = i\pi e^{ia\sigma}.
\ee

From the result (\ref{xr}) we have that 
\bea
\int\limits_{-\infty}^{+\infty} dx \frac{x\cos(ax)}{(x^2-\sigma^2)} & = & 0 \label{cos}\\
\int\limits_{-\infty}^{+\infty} dx \frac{x\sin(ax)}{(x^2-\sigma^2)} & = & \pi\,e^{ia\sigma} \label {sin}.
\eea

Our original quantum mechanics integral $S(\sigma)$ is our result (\ref{sin}) for $a=1$, so
\bea
E_x^+(1,\sigma) & \equiv & S(\sigma)\nonumber\\
                           & = & \int\limits_{-\infty}^{+\infty} dx \frac{x\sin(x)}{(x^2-\sigma^2)}  =  \pi\,e^{i\sigma}.
\eea

The above result for quantum mechanical scattering problem corresponds to an outgoing wave. This result is equivalent to letting $\sigma \to \sigma + i\varepsilon$ in the Cauchy contour technique calculation. Noting that the integral on the left handside is invariant under $\sigma \leftrightarrow -\sigma$, we have the result for quantum mechanical scattering problem corresponding to an incoming wave.
\bea
E_x^-(1,-\sigma) & \equiv & S(-\sigma)\nonumber\\
                            & = & \int\limits_{-\infty}^{+\infty} dx \frac{x\sin(x)}{(x^2-\sigma^2)}  =  \pi\,e^{-i\sigma}.
\eea

This corresponds to letting $\sigma \to \sigma - i\varepsilon$ in the Cauchy residue calculation. And the PV result for the integral is, as before, 
\bea
S_{\rm PV} (\sigma) & = &  \frac{1}{2}\left\{ E_x^+(1,\sigma)+E_x^-(1,-\sigma)\right\} =  \frac{1}{2}\left\{ S(\sigma)+S(-\sigma)\right\}\nonumber\\
                                & = &  \pi\left\{\frac{e^{i\sigma}+e^{-i\sigma}}{2}\right\} = \pi \cos \sigma.
\eea

Since the original integral $S(\pm \sigma)$ is an improper integral, we can see that the value we ascribe to it will depend on which type of physical boundary conditions we want: incoming or outgoing waves or even the average between those boundary conditions (PV).  

\section{Conclusions}

We have shown that using the NDIM technique, it was possible to evaluate some improper integrals and more specifically, the one integral that is relevant for quantum mechanical scattering problems. We have shown how to do the calculations in the NDIM technique and shown that the results it provides are equivalent to the results obtained using the Cauchy residue procedure, in which poles in the real axis is given an infinitesimal shift into the complex plane, either $\sigma + i\varepsilon$ or $\sigma - i\varepsilon$. In the NDIM procedure, each pole residue gives a distinct answer \cite{Suzuki}. In our present calculation for the quantum mechanical scattering problem, there are two poles, and so two residues; and the PV result can be calculated by the simple average between the two residue calculations.


\end{document}